\documentclass[preprint, pre,superscriptaddress, letterpaper, fleqn, floatfix, showpacs ]{revtex4}
\usepackage{euscript, units, amssymb,amsfonts,amsmath, graphics, graphicx, dcolumn, fancyhdr}

\usepackage{hyperref}

\begin{document}
\title{Wave spectra of 2D dusty plasma solids and liquids}
\author{Lu-Jing Hou}
\affiliation{IEAP, Christian-Albrechts Universit\"{a}t, Kiel, Germany}
\author{Z.\ L.\ Mi\v{s}kovi\'{c}}
\affiliation{Department of Applied Mathematics, University of Waterloo, Waterloo, Ontario, Canada N2L 3G1}
\author{Alexander Piel}
\affiliation{IEAP, Christian-Albrechts Universit\"{a}t, Kiel, Germany}
\author{Michael S. Murillo}
\affiliation{Physics Division, Los Alamos National Laboratory, Los Alamos, NM 87545, USA}

\date{\today}
\begin{abstract}
Brownian dynamics simulations were carried out to study wave spectra of two-dimensional dusty plasma liquids and solids for a
wide range of wavelengths. The existence of a longitudinal dust thermal mode was confirmed in simulations, and a cutoff
wavenumber in the transverse mode was measured. Dispersion relations, resulting from simulations, were compared with those from
analytical theories, such as the random-phase approximation (RPA), quasi-localized charged approximation (QLCA), and harmonic
approximation (HA). An overall good agreement between the QLCA and simulations was found for wide ranges of states and
wavelengths after taking into account the direct thermal effect in the QLCA, while for the RPA and HA good agreement with
simulations were found in the high and low temperature limits, respectively.
\end{abstract}
\maketitle

\section{Introduction}
A laboratory-generated dusty plasma is a suspension of micron-sized particles immersed in a usual plasma with ions, electrons and
neutral gas molecules \cite{Shukla2001,Shukla2002,Shukla2008}. Dust particles acquire a few thousand of electron charges by
absorbing the surrounding electrons and ions, and they consequently interact with each other via a dynamically-screened Coulomb
potential \cite{Lampe2000}, while undergoing Brownian motions due to frequent collisions, mainly with the neutral gas molecules. When the
interaction potential energy between charged dust particles significantly exceeds their kinetic energy, they become strongly
coupled and, consequently, they can form ordered structures characteristic of a liquid or solid state. Such structures are
commonly referred to as strongly coupled dusty plasmas (SCDPs).

Two-dimensional (2D) SCDPs have become particularly favored in recent laboratory experiments, because in 2D geometry the complication due to ion wake effect (see, e. g., \cite{Lampe2000} and references therein) should be absent and the particle interaction can be well approximated by the Yukawa potential \cite{Lemons2000,Konopka2000}. Of particular
interest in 2D dusty plasmas are their collective and dynamical properties, such as longitudinal and transverse wave modes,
which have been studied extensively over the past decade in experiments
\cite{Homann1997,Homann1998,Nunomura2002,Zhdanov2003,Nunomura2005,Nosenko2006}, theories
\cite{Peeters1987,Melandso1996,Dubin2000,Wang2001,Murillo2003,Kalman2004,Piel2006}, and numerical simulations
\cite{Kalman2004,Sullivan2006,Hartmann2007}. On the experimental side, externally excited (longitudinal) dust lattice waves
(DLWs) were first observed by Homann \emph{et al.} \cite{Homann1997,Homann1998}, who found their dispersion relation to be in
good agreement with the theoretical prediction of Melands\o \, \cite{Melandso1996}. Next, thermally excited phonon spectrum in a
2D plasma crystal was observed in an experiment by Nunomura \emph{et al.} \cite{Nunomura2002}, who demonstrated a good agreement
of this spectrum with theory \cite{Peeters1987,Dubin2000,Wang2001} for both the longitudinal and transverse modes in the entire
first Brillouin zone. That experiment was later extended by Zhdanov \emph{et al.} \cite{Zhdanov2003}, who studied the
polarization of wave modes in a 2D plasma crystal, and by Nunomura \emph{et al.} \cite{Nunomura2005}, who studied the wave
spectra in both liquid and solid states for a wide range of wavelengths. More recently, Nosenko \emph{et al.} \cite{Nosenko2006}
have measured experimentally a critical cutoff wave number for shear waves in 2D dusty plasma liquids.

On the theoretical side, besides the above mentioned DLW theories of Peeters and Wu \cite{Peeters1987}, Melands\o
\cite{Melandso1996},\, Dubin \cite{Dubin2000}, and Wang \emph{et al.} \cite{Wang2001}, as well as the standard dust acoustic
wave (DAW) theory \cite{Rao1990,Rao2000} along with its 2D derivatives \cite{Stenflo2000,Hou2004}, there have been many other studies of SCDPs. For example, a semi-analytic approximation was used to study wave propagation in 2D SCDPs \cite{Dubin2000,Hou2004}. The quasi-localized charge approximation (QLCA) \cite{Golden2000} was used to study collective modes and dynamics of both 3D \cite{Rosenberg1997,Kalman2000} and 2D \cite{Kalman2004} dusty plasma liquids. Furthermore, a generalized hydrodynamic (GH) model was adopted by Kaw and Sen \cite{Kaw1998} to study wave dispersion in 3D dusty plasma liquids, while Murillo and coworkers studied collective modes by using both kinetic \cite{Murillo1998} and the GH \cite{Murillo2000,Murillo2000pop} methods. The latter method was used, in particular, to study critical wavenumbers for
transverse modes in a 3D dusty plasma in liquid phase \cite{Murillo2000}. On the other hand, computer simulations
\cite{Winske1999,Ohta2000,Kalman2004,Sullivan2006,Hartmann2007}, as an essential supplement to real experiments, have played
important roles in validating various analytical theories and in explaining experimental observations. Both Molecular dynamics
(MD) and the particle-in-cell (PIC) simulations were carried out by Winske \emph{et al.} \cite{Winske1999} to study longitudinal
wave dispersion in one-dimensional systems. Results were compared with the DAW mode and with the 3D QLCA \cite{Rosenberg1997}
including the strong-coupling effects, and an agreement with the QLCA was found only at very long wavelengths due to the small
system size. Later on, Ohta and Hamaguchi \cite{Ohta2000} studied both longitudinal and transverse dispersion relations in 3D
dusty plasma liquids, and results were compared with both the QLCA \cite{Rosenberg1997,Kalman2000} and the GH model
\cite{Kaw1998,Murillo2000}. It was found that, for the longitudinal mode, both theories were in close agreement with simulations
in the entire first Brillouin zone, whereas, for the transverse mode, good agreements with simulations were found for both
theories, except in the very long wavelength region, where the GH model successfully predicts a cutoff but the QLCA does not.
This deficiency of the QLCA can be rectified by introducing a phenomenological damping which accounts for the diffusional and
other damping effects, as was shown more recently in the work by Kalman \emph{et al.} \cite{Kalman2004}, where the QLCA was
extended to study 2D dusty plasma liquids, and the results were critically compared with MD simulations. Good agreements with
simulations were found for both the longitudinal and transverse modes, mostly in the first Brillouin zone. (Please see Ref. \cite{Donko2008} for a nice review of recent development about collective dynamics in 2D and 3D Yukawa liquids, based mainly on QLCA and computer simulations.)

However, it should be noted that, first, most of the above simulations
\cite{Winske1999,Ohta2000,Kalman2004,Sullivan2006,Hartmann2007,Donko2008} considered dusty plasmas in the liquid state, and the obtained
dispersion relations were limited to the first Brillouin zone. To the best of our knowledge, there are no simulations verifying
the above analytical theories for a wider range of wavelengths, as well as for dusty plasmas in non-ideal gaseous, or in solid
states. Secondly, no simulations have been conducted to show a transition from the random phase approximation (RPA) based DAW to
the harmonic approximation (HA) based DLW when dusty plasma goes from a high temperature liquid (or non-ideal gaseous) state to
a low temperature crystalline state. Third, an analytical theory based on the RPA \cite{Rao2000} predicted a so-called dust
thermal wave (DTW) due to the direct thermal effect of dust, and its existence was verified in a subsequent experiment
\cite{Nunomura2005}. However, the original theory \cite{Rao2000} applies only to weakly coupled systems, and it would be
interesting to study, both analytically and via simulation, how the DTW behaves in strongly coupled systems.

To fulfill these goals, we perform here computer simulations and study the resulting wave spectra of 2D dusty plasmas for a wide
range of wavelengths and system states. Results are compared with the available analytical theories for the 2D geometry, such
as, RPA, QLCA and HA. In addition, two extensions of the standard QLCA are discussed, one including the direct thermal effect,
and the other implementing a critical wave cutoff in the transverse mode, based on the method of Kalman \emph{et al.}
\cite{Kalman2004,Donko2008}. The role of damping effects on collective modes is also investigated in the simulation. The remaining part of
the paper is organized as follows. Details of our simulations are presented in Sec.\ II, which is followed by a presentation and
discussion of the results in Sec.\ III, including brief derivations and reviews of various analytical theories. Concluding
remarks are given in Sec. IV.

\section{Simulation}
\subsection{Algorithm}
Our simulation is based on the Brownian dynamics (BD) method \cite{Allen1989,BD2008,BDM2009}, which may be regarded as a generalization
of the standard MD method. Namely, while the MD simulation is based on Newton's equations of motion, the BD method is based on
their generalization in the form of Langevin equation (and its integral), viz.,
\begin{eqnarray}
\frac{d}{dt}\mathbf{r}&=&\mathbf{v} \nonumber \\
\frac{d}{dt}\mathbf{v}&=&-\nu \mathbf{v}+\frac{1}{m}\mathbf{F}+\mathbf{A}(t) , \label{eqlangevin}
\end{eqnarray}
where, as usual, $m$, $\mathbf{v}$ and $\mathbf{r}$ are, respectively, the mass, velocity and position of a Brownian particle,
and $\mathbf{F}$ is the systematic (deterministic) force coming from the inter-particle interactions within the system and,
possibly, from external force fields. What is different from Newton's equations is the appearance of dynamic friction, $-\nu
\mathbf{v}$, and the Brownian acceleration, $\mathbf{A}(t)$, which represent complementing effects of a single, sub-scale
phenomenon: numerous, frequent collisions of Brownian particle with molecules in the medium. While the former represents the
average effect of these collisions, the latter represents fluctuations due to discreteness of the collisions, and is generally
assumed to be well represented by a delta-correlated Gaussian white noise. They are both related to the medium temperature
through a fluctuation-dissipation theorem.

The Langevin equations Eq.\ (\ref{eqlangevin}) may be numerically integrated in a manner similar to the integration of Newton's
equations in the MD method, and such a technique is generally called Brownian dynamics. (Note that several different names are
used in literature to designate methods for numerical integration of Langevin equation, for example, Brownian Dynamics, Langevin
Dynamics, or Langevin Molecular Dynamics, depending on the background, area, or preferences of different researchers. Those
names may refer to the same technique that we discuss here, or they may involve subtle differences in derivations of the
simulation formulae, and/or in implementations of simulation. We follow here definitions given by Allen and Tildesley
\cite{Allen1989}.) The advantages of BD here are as follows. Firstly it brings the simulation closer to real dusty plasma experiments by taking into account both the Brownian motion and damping effect self-consistently. And secondly it simplifies the simulation in a sense that no external thermostat is need.

We employ here the 5th order Gear-like predictor-corrector algorithm \cite{BD2008,BDM2009} for BD simulation, which was used
successfully in simulating shock wave propagation \cite{Jiang2006,Hou2006,Hou2008,Hou2008pop}, heat conduction \cite{SCCS2008},
and diffusion processes in SCDPs \cite{Hou2009}. When compared with several other popular methods for BD simulation, such as
Euler-like, Beeman-like and Verlet-like methods \cite{Allen1989}, the present method can cover a wider range of friction
coefficients $\nu$, and is particularly reliable in the low-damping regime while exhibiting higher-order accuracy, good
stability, and negligible drift on long time scales. Therefore, the present algorithm appears to be especially suitable to
simulate dusty plasmas, which are often slightly damped.

\subsection{Calculation of wave spectra}
\begin{figure}[htp]
\centering
\includegraphics[trim=15mm 5mm 12mm 10mm,clip, width=0.7\textwidth]{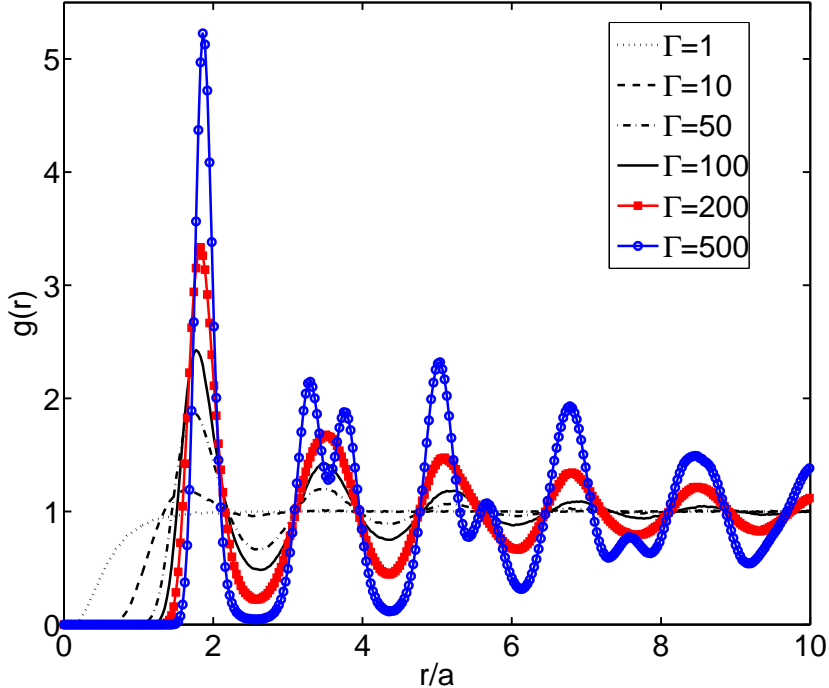}
\caption{(Color online) Radial distribution function $g(r)$ for $\kappa=1$ and different $\Gamma$.} \label{fig_gr_k1}
\end{figure}

Typically, $N=4000$ dust particles are simulated in a square with periodic boundary conditions. Particles interact with each
other via pairwise Yukawa potentials \cite{Konopka2000} of the form $\phi(r)=(e^2Z^2_{d}/r)\exp{(-r/\lambda_{D})}$, where
$e$ is the elementary charge, $Z_{d}$ is the average number of charges on each dust particle, $r$ the inter-particle distance,
and $\lambda_{D}$ the Debye screening length. Note here that the Yukawa interaction is not merely an assumption, but is rather
consistent with theoretical derivations, described in the following section. Such a system can be fully characterized by three
parameters \cite{Fortov2003}: the Coulomb coupling parameter, $\Gamma=(eZ_{d})^2/(aT)$, the screening parameter,
$\kappa=a/\lambda_{D}$, and the damping rate, $\nu/\omega_{pd}$, due to the neutral gas, where $T$ is the system temperature (in energy
units), $a=(\pi \sigma_{d0})^{-1/2}$ the Wigner-Seitz radius, and $\sigma_{d0}$ the equilibrium dust number density. The dust
plasma frequency is defined by $\omega_{pd}=\left[2(eZ_{d})^{2}/(m a^{3})\right]^{1/2}$, where $m$ is the dust particle mass. We
shall use $\kappa=1$ throughout our simulation and in subsequent discussions, because $\kappa\sim 1$ is the most typical value
found in experiments and, moreover, no substantial effects of varying $\kappa$ are expected in the features to be discussed in
the following.

Initially, dust particles are randomly placed in the square. The system comes to an equilibrium after a period which is
proportional to $1/\nu$, and is usually about $10/\nu$. The time step used in simulations is $\delta t=0.02\omega^{-1}_{pd}$ for
$\Gamma\geq 10$ and $\delta t=0.01\omega^{-1}_{pd}$ for $\Gamma<10$. After the system reaches an equilibrium, radial
distribution function $g(r)$ and wave spectra are calculated.

Figure \ref{fig_gr_k1} shows examples of $g(r)$ for $\kappa=1$, with different $\Gamma$ values. These results will be used as an
input below in evaluating the QLCA dispersion relations.

The longitudinal, $\mathcal{L}(\mathbf{k},\omega)$, and transverse, $\mathcal{T}(\mathbf{k},\omega)$, wave spectra are
determined by means of the current-current correlation functions in the longitudinal and transverse directions, respectively. We
follow Ref.\ \cite{Boon1980}, in which the current density of the system is defined by
\begin{equation}
\mathbf{j}(\mathbf{r},t)=\frac{1}{\sqrt{N}}\sum^{N}_{i=1}\mathbf{v}_{i}(t)\delta[\mathbf{r}-\mathbf{r}_{i}(t)] ,
\label{Eq_current_density_t}
\end{equation}
with the Fourier transform of its cartesian component $\alpha$ given by
\begin{equation}
j_{\alpha}(\mathbf{k},t)=\frac{1}{\sqrt{N}}\sum^{N}_{i=1}v_{i\alpha}(t)e^{i\mathbf{k}\cdot\mathbf{r}_{i}(t)} ,
\label{Eq_current_density_kt}
\end{equation}
where $\mathbf{r}_{i}(t)$ and $\mathbf{v}_{i}(t)$ are, respectively, position and velocity of the $i$th particle at time $t$,
with $\alpha$ being $x$ or $y$. Assuming that waves propagate along the $x$ direction, i.e., $\mathbf{k}=\{k,0\}$, one defines
\begin{eqnarray}
J_{L}(\mathbf{k},t)=\langle j^{*}_{x}(\mathbf{k},t)j_{x}(\mathbf{k},0) \rangle \nonumber, \\
J_{T}(\mathbf{k},t)=\langle j^{*}_{y}(\mathbf{k},t)j_{y}(\mathbf{k},0) \rangle \nonumber,
\end{eqnarray}
to be, respectively, longitudinal and transverse current auto-correlations. Here, the asterisk designates complex conjugation,
while the angular brackets indicate an ensemble average. The longitudinal and transverse wave spectra are then obtained by
Fourier transforms of the corresponding current auto-correlations \cite{Boon1980}
\begin{eqnarray}
\mathcal{L}(\mathbf{k},\omega)=\int^{+\infty}_{-\infty}{dt\ e^{i\omega t}J_{L}(\mathbf{k},t)}, \nonumber \\
\mathcal{T}(\mathbf{k},\omega)=\int^{+\infty}_{-\infty}{dt\ e^{i\omega t}J_{T}(\mathbf{k},t)} . \label{Eq_spectra}
\end{eqnarray}
In the simulation, Eqs.\ (\ref{Eq_spectra}) are evaluated by using discrete Fourier transform in a period of $655
\omega^{-1}_{pd}$, and the results are averaged over a longer period of $10480 \omega^{-1}_{pd}$.

\section{Results and discussions}
In this section, we present our main simulation results for wave spectra, together with analytical results for dispersion
relations for DAW \cite{Hou2004}, DTW \cite{Rao2000}, and those obtained by using QLCA \cite{Kalman2004} and HA
\cite{Dubin2000}. The relevant theoretical derivations are briefly reviewed for the sake of completeness.

\subsection{DAW}
Let us begin with a weakly coupled state. We use here a fluid description for a mono-layer of dust particles levitating in
plasma, without considering details of their mutual interactions. Assuming that a cold, 2D dust fluid occupies the plane $z=0$
in a Cartesian coordinate system with ${\bf R}=\{x,y,z\}$, we let $\sigma _d({\bf r},t)$ and ${\bf u}_d({\bf r},t)$ be,
respectively, number density per unit area and velocity field (having only the $x$ and $y$ components) of the dust fluid at the
position ${\bf r}=\{x,y\}$ and at time $t$. The continuity equation and the momentum equation for the fluid are, respectively
\cite{Hou2004}
\begin{equation}
\frac{\partial \sigma _d({\bf r,}t)}{\partial t}+\nabla _{\parallel }\cdot \left[ \sigma _d({\bf r,}t){\bf u}_d({\bf
r,}t)\right] =0 ,  \label{Eq_continuity}
\end{equation}
\begin{equation}
\frac{\partial {\bf u}_d({\bf r,}t)}{\partial t}+{\bf u}_d({\bf r,}t)\cdot \nabla _{\parallel }{\bf u}_d({\bf
r,}t)=\frac{eZ_d}{m}\nabla _{\parallel }\Phi ({\bf R},t)\Big\vert_{z=0}-\nu {\bf u}_d({\bf r,}t)\qquad ,  \label{Eq_momentum}
\end{equation}
where $\nu $ is the Epstein drag coefficient. Note that the spatial differentiation in Eqs.\ (\ref{Eq_continuity}) and
(\ref{Eq_momentum}) only includes tangential directions, viz., $\nabla _{\parallel }=\displaystyle{\hat{{\bf x}}\frac
\partial {\partial x}+\hat{{\bf y}}\frac \partial {\partial y}}$. The first term in the right-hand side of Eq.\
(\ref{Eq_momentum}) indicates that, although the total electrostatic potential $\Phi({\bf R},t)$ depends on all three spatial
coordinates ${\bf R}\equiv\{{\bf r},z\}$, only the $x$ and $y$ components of the electrostatic force, evaluated in the plane
$z=0$, affect the motion of the dust fluid. The full spatial dependence of the electrostatic potential $\Phi$ is determined by
the Poisson equation in 3D,
\begin{equation}
\quad \nabla ^2\Phi ({\bf R},t)=-4\pi e\left[ n_i({\bf R,}t)-n_e({\bf R,} t)-Z_d\sigma _d({\bf r,}t)\,\delta (z)\right] \qquad ,
\label{Eq_Poisson}
\end{equation}
where $\nabla =\displaystyle{\ \nabla _{\parallel }+\hat{{\bf z}}\frac
\partial {\partial z}}$. The electron and ion volume densities are given by
Boltzmann relations, $n_e=n_0\exp (e\Phi /T_e)$ and $n_i=n_0\exp (-e\Phi /T_i)$, respectively, owing to the fact that the
dynamics of massive dust particles is so slow that both electrons and ions are considered to have enough time to reach their
respective local equilibria, with $n_{0}$ being the equilibrium plasma density and $T_{i(e)}$ the ion (electron) temperature (in
energy units).

The above equations can be solved perturbatively \cite{Piel2006,Hou2004}, giving a dielectric function of the 2D dust fluid in the form
\begin{equation}
\varepsilon (k,\omega )=1-\frac{2\pi e^2Z_d^2\sigma _{d0}}{m\lambda _D}\frac{k^2\lambda _D^2}{\sqrt{k^2\lambda _D^2+1}}
\frac{1}{\omega (\omega +i\nu )} . \label{Eq_dielectric}
\end{equation}
We define
\begin{equation}
\omega_{0}^{2}(k) =\frac{2\pi e^2Z_d^2\sigma _{d0}}{m\lambda _D}\frac{(\lambda_{D}\,k)^{2}}{\sqrt{k^2\lambda
_D^2+1}}\equiv\frac{\sigma_{d0}}{m}\widetilde{\phi}(k)k^{2} ,  \label{Eq_omega0}
\end{equation}
and note that $\widetilde{\phi}(k)$ is the 2D Fourier transform of the Yukawa potential. A dispersion relation for acoustic waves in the 2D dust fluid \cite{Hou2004} can be obtained from
\begin{equation}
\omega(\omega+i\nu) =\omega_{0}^{2}(k)  ,  \label{Eq_dispersion_RPA}
\end{equation}
in analogy to the 3D case \cite{Rao1990,Rao2000,Shukla2001,Shukla2002}. The real part of the dispersion is shown in Fig. \ref{fig_RPA_DTW} (heavy dashed lines), while the corresponding discussion is postponed to the following subsection.

\subsection{Dust thermal wave (DTW)}
\begin{figure}[htp]
\centering
\includegraphics[trim=5mm 0mm 12mm 0mm,clip, width=0.6\textwidth]{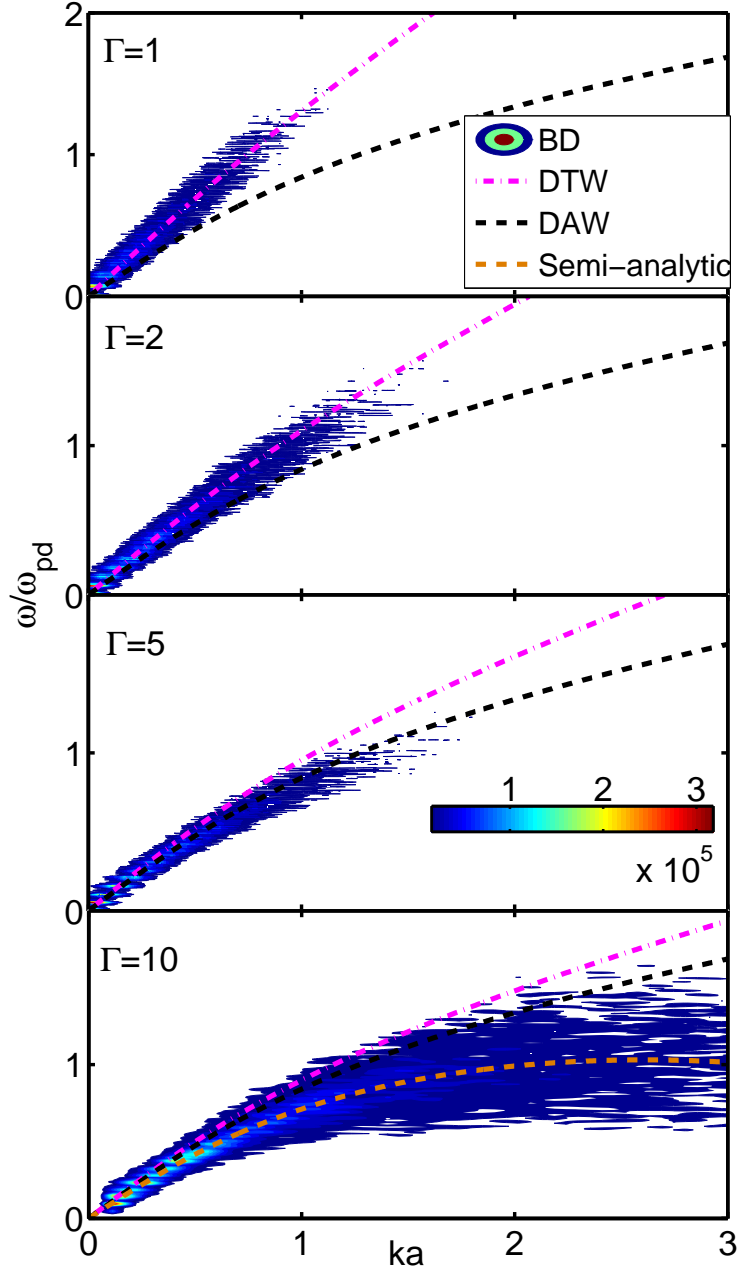}
\caption{(Color online) (Color bar in arbitrary unit) Wave spectra of longitudinal mode for $\kappa=1$, $\nu=0.01\omega_{pd}$ and different $\Gamma$ (nonideal gas states).
Dashed lines are dispersion relations of DAW and result of semi-analytic approximation, respectively, while dash-dotted lines are results of DTW.} \label{fig_RPA_DTW}
\end{figure}

Note that, in the above derivation, we have assumed that the dust fluid is cold and, consequently, we neglected the direct
thermal effect. Nevertheless, it would be interesting to see how this effect changes wave dispersion in the dust fluid. This
effects can be easily retained by including the ideal-gas part of pressure $P_{dk}$ in the momentum balance equation, Eq.\
(\ref{Eq_momentum}), as follows \cite{Rao2000}
\begin{equation}
\frac{\partial {\bf u}_d({\bf r,}t)}{\partial t}+{\bf u}_d({\bf r,}t)\cdot \nabla _{\parallel }{\bf u}_d({\bf
r,}t)=\frac{eZ_d}{m}\nabla _{\parallel }\Phi ({\bf R},t)\Big\vert_{z=0}-\nu {\bf u}_d({\bf
r,}t)+\frac{1}{m\sigma_{d0}}\nabla_{\parallel}P_{dk},  \label{Eq_momentumn}
\end{equation}
where $P_{dk}=\gamma_{d} k_{B}T_{d}\sigma_{d}$ is the kinetic part of the pressure in dust component, and $\gamma_{d}=2$ is the adiabatic index for the 2D dust system \cite{Rao2000}.

With this new momentum balance equation, the dielectric function of the 2D dust fluid becomes
\begin{equation}
\varepsilon (k,\omega )=1- \frac{1}{\omega (\omega +i\nu )} \left [\omega_{0}^2(k) +\frac{\gamma_{d}v^{2}_{th}}{2}k^{2} \right
], \label{Eq_dielectric_DTW}
\end{equation}
with $v_{th}=\sqrt{2k_{B}T_{d}/m}$ being the thermal speed of dust particles, which gives the dispersion relation for DTW
\begin{equation}
\omega(\omega+i\nu) =\omega^{2}_{0}(k)+\frac{\gamma_{d}v^{2}_{th}}{2}k^{2} .  \label{Eq_dispersion_DTW}
\end{equation}
The real part of the DTW dispersion is shown in Fig. \ref{fig_RPA_DTW} (dash-dotted lines).

One expects that the DAW and DTW modes, derived above, should be dominant in weakly coupled systems. However, to the best of our
knowledge, this conjecture has not been examined in previous simulations. To elucidate this issue, we show in Fig.\
\ref{fig_RPA_DTW} wave spectra from simulations, together with the dispersion relations for both DAW and DTW for very low
coupling strengths, at which systems are often regarded as being in a non-ideal gaseous state. One sees that, at such high
temperatures, the collective modes are heavily damped at short wavelengths. There are essentially no collective modes beyond
$ka=2$ for $\Gamma=1$, $2$ and $5$ and, moreover, these modes diminish at higher temperatures, or lower $\Gamma$s. One would attribute this to Landau
damping and viscous/collisional damping. Comparison with the above analytical results at $\Gamma=1$ shows that the agreement between the DTW and the simulation
is remarkably good, whereas the DAW displays noticeable discrepancy with simulation, indicating that the thermal effect is
significant. With increasing $\Gamma$ or, equivalently, decreasing temperature, a discrepancy between the DTW and simulation
develops and becomes noticeable at, e.g., $\Gamma=5$, whereas the DAW seems to agree better with simulation. As we'll see in the next subsection, this seemingly better agreement between DAW and simulation at $\Gamma=5$ is just a coincidence, because the direct thermal effect and the strong coupling effect cancel each other, as will be discussed in the following subsection. The discrepancy arises immediately when $\Gamma$ further increases, say at $\Gamma=10$.

\subsection{Semi-analytic approximation}
The DAW and DTW derived in the previous subsections are essentially based on the mean-field theory of RPA \cite{Rao1990,Rao2000}, in which the short-range interactions between dust particles are neglected. Therefore, those results are applicable, in principle,  only in weakly coupled situations, i.e., when $\Gamma<<1$. For strongly coupled systems, the short-range inter-particle interaction becomes important and it is necessary to take this effect into account. Following the line of reasoning in our fluid description above, this can be done immediately by introducing the interaction part of the pressure $P_{di}$ into the momentum balance equation \cite{Hou2004, Ichimaru}
\begin{equation}
\frac{\partial {\bf u}_d({\bf r,}t)}{\partial t}+{\bf u}_d({\bf r,}t)\cdot \nabla _{\parallel }{\bf u}_d({\bf
r,}t)=\frac{eZ_d}{m}\nabla _{\parallel }\Phi ({\bf R},t)\Big\vert_{z=0}-\nu {\bf u}_d({\bf
r,}t)+\frac{1}{m\sigma_{d0}}\nabla_{\parallel}P_{dk}+\frac{1}{m\sigma_{d0}}\nabla_{\parallel}P_{di},  \label{Eq_momentumn1}
\end{equation}
where $P_{di}$ usually contains information of system structure and interaction. A semi-analytic results of dielectric function \cite{Dubin2000,Hou2004} can then be obtained by relating $P_{di}$ to correlation energy $\epsilon_{c}$ through density functional theory \cite{Ichimaru}, i, e., 
\begin{equation}
\nabla_{\parallel}P_{di}=\left ( \frac{\delta P_{di}}{\delta \sigma_{d}} \right )_{T}\nabla_{\parallel}\sigma_{d}.
\end{equation}
Here one sees that $\alpha=(\delta P_{di}/\delta \sigma_{d}){T}$ is actually the isothermal compressibility \cite{Kaw1998,Ichimaru} and can be further written as \cite{Hou2004}
\begin{equation}
\alpha=\frac{\sigma_{d0}}{m}\frac{\partial^{2}}{\partial \sigma^2_{d0}}[\sigma_{d0}\epsilon_{c}(\sigma_{d0})]. \label{Eq_alpha}
\end{equation}
(See Ref. \cite{Hou2004} and references therein for more details of the derivation.) Empirical expression of $\epsilon_{c}$ for 2D Yukawa system is now available through computer simulation \cite{Hartmann2005}. One has \cite{Dubin2000,Hou2004}
\begin{equation}
\varepsilon\left( k,\omega \right) =1-\frac{1}{\omega(\omega+i\nu)-\alpha k^2}\left [\omega_{0}^2(k) +\frac{\gamma_{d}v^{2}_{th}}{2}k^{2} \right
], \label{Eq_dielectric_SA}
\end{equation}
and consequently the dispersion relation
\begin{equation}
\omega(\omega+i\nu) =\omega^{2}_{0}(k)+\frac{\gamma_{d}v^{2}_{th}}{2}k^{2} -\alpha k^{2}.  \label{Eq_dispersion_SA}
\end{equation}
Here the term with $\alpha$ in Eq. (\ref{Eq_dielectric_SA}) can be regarded as a local field correction in the mean-field theory accounting for the correction due to the strong coupling effect \cite{Ichimaru}. Because $\alpha$ is related to the correlation energy $\epsilon_{c}$, which in turn can be obtained from simulation \cite{Hartmann2005}, this dispersion relation was entitled as a semi-analytic approximation \cite{Dubin2000}. Typical value of $\alpha$ is about $0.2 \sim 0.3\omega_{pd}a$.

It should be noted that in Eq. (\ref{Eq_dispersion_SA}), the direct thermal effect and the correlation effect have opposite influence on the dispersion relation. At very high temperature, say $\Gamma=1$ in Fig. \ref{fig_RPA_DTW}, the correlation effect is negligibly small, so we saw a perfect agreement between DTW and simulation. With the increase of $\Gamma$, the former becomes smaller and smaller while the latter becomes more and more pronounced and at $\Gamma=5$ the two effects somehow cancel each other, so we saw a good agreement between DAW and simulation. As can be expected, with the further increase of $\Gamma$, the correlation effect becomes dominant, so that the semi-analytic result gives a good agreement with simulation at $\Gamma=10$ in Fig. \ref{fig_RPA_DTW}. The similar effect had also been observed in GH theory of Kaw and Sen \cite{Kaw1998} for 3D dusty plasma liquids.

\subsection{QLCA}
\begin{figure}[htp]
\centering
\includegraphics[trim=15mm 10mm 12mm 10mm,clip, width=0.7\textwidth]{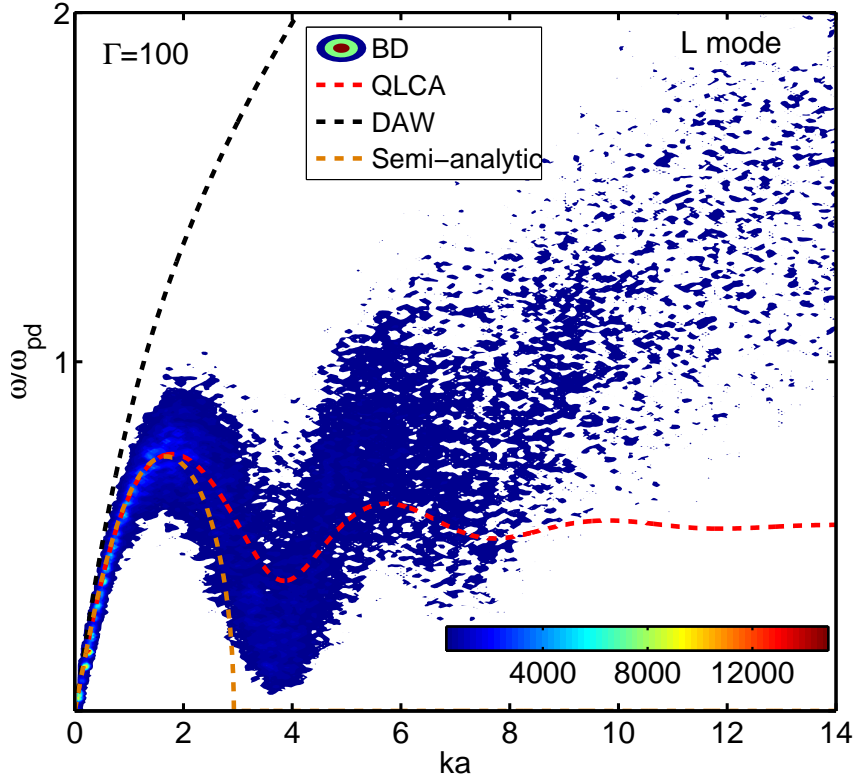}
\caption{(Color online) (Color bar in arbitrary unit) Wave spectra of Longitudinal mode for $\kappa=1$, $\Gamma=100$ and $\nu=0.01\omega_{pd}$. Dashed lines are dispersion
relations of DAW and QLCA and semi-analytic approximation.} \label{fig_LT_K1G100}
\end{figure}
\begin{figure}[htp]
\centering
\includegraphics[trim=5mm 4mm 10mm 0mm,clip, width=0.7\textwidth]{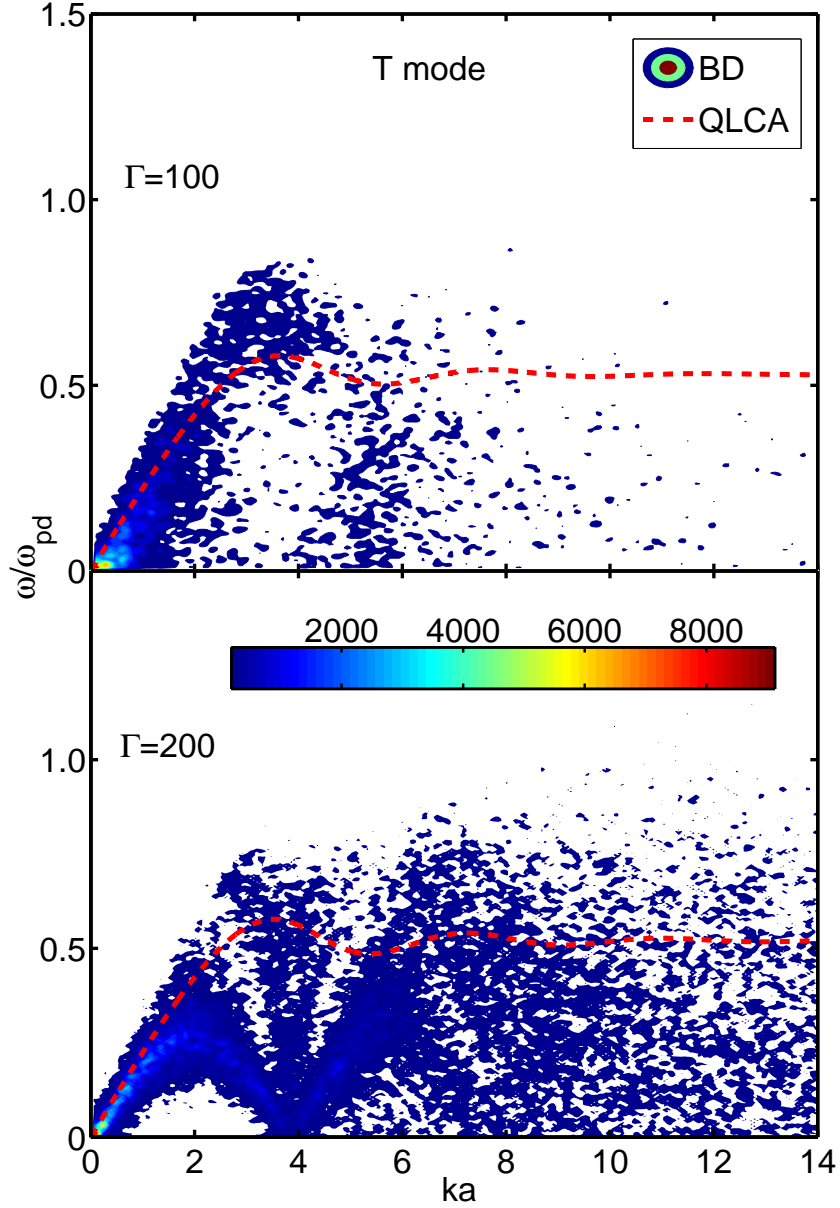}
\caption{(Color online) (Color bar in arbitrary unit) Wave spectra of transverse mode for $\Gamma=100$ and $200$, $\kappa=1$ and $\nu=0.01\omega_{pd}$. Dashed lines are
dispersion relation of QLCA.} \label{fig_T_K1G100}
\end{figure}

In previous subsection, we had come up to the concept of local field correction (LFC) and introduced the LFC through density functionals of interaction part of pressure, i. e., Eq. (\ref{Eq_alpha}). In the following, we adopt the LFC in terms of the QLCA \cite{Golden2000,Kalman2004,Donko2008}, which had been found quite successful in describing collective dynamics in strongly coupled dusty plasmas \cite{Kalman2004,Piel2006,Donko2008}. The corresponding dielectric function (for both longitudinal and transverse modes) of a strongly coupled 2D dusty plasma reads as follows
\begin{equation}
\varepsilon_{L/T}\left( k,\omega \right) =1-\frac{\omega_0^2(k)}{\omega(\omega+i\nu)-D_{L/T}(k)}, \label{Eq_dielectric_QLCA}
\end{equation}
where $D_L(k)$ and $D_T(k)$ are projections of the dynamical matrix in the longitudinal and transverse directions, respectively,
and are functionals of the equilibrium radial distribution function, $g(r)$. Detailed expressions for $D_L(k)$ and $D_T(k)$ can
be found in Refs.\ \cite{Kalman2004,Jiang2006}. Note here that the direct thermal effect is neglected in QLCA and will be resumed in next subsection.

The longitudinal and transverse modes are determined from these dielectric functions by letting,
\begin{equation}
\varepsilon_L\left( k,\omega \right)=0,\,\,\, \text{and} \,\,\, \varepsilon_T^{-1}\left( k,\omega \right)=0, \label{Eq_disp}
\end{equation}
which give
\begin{eqnarray}
\omega(\omega+i\nu)&=& \omega_0^2(k)+D_L(k)  \label{Eq_disp_L}  \\
\omega(\omega+i\nu)&=& D_T(k), \label{Eq_disp_T}
\end{eqnarray}
for the longitudinal and transverse dispersion relations, respectively.

Figure \ref{fig_LT_K1G100} shows a longitudinal phonon spectrum for $\Gamma=100$ and $\nu=0.01\omega_{pd}$. Also shown are the dispersion relations of DAW, semi-analytic approximation and QLCA. A good agreement between the simulation and the QLCA is clearly seen in a broad range of wavelength on the long wavelength side, as expected \cite{Kalman2004,Piel2006}. Discrepancy appears beyond the first Brillouin zone, especially for very large wave numbers, where the simulation displays a raising tail due to the DTW mode, while the QLCA converges to the Einstein frequency \cite{Golden2000,Kalman2004}. This discrepancy is a consequence of the neglect of the direct thermal effects in the QLCA, which prompts us to discuss a suitable amendment to this theory, to be described in the following subsection. It is not surprising to find that the result of semi-analytic approximation is essentially the same as QLCA in the long wavelength region (up to around $ka=2.0$), because actually $\alpha=\lim_{k\rightarrow 0}D_{L}(k)/k^2$ \cite{Golden2000,Kalman2004,Donko2008}. However, the semi-analytic approximation breaks down for short wavelength.

Figure \ref{fig_T_K1G100} shows the wave spectra of a transverse mode with $\nu=0.01\omega_{pd}$, for $\Gamma=100$ and $200$,
along with a dispersion relation from the QLCA. The transverse mode is heavily damped, especially at short wavelengths and in
the system at a higher temperature, as can be seen from Fig.\ \ref{fig_T_K1G100}. For $\Gamma=100$, there are practically no
transverse modes beyond $ka\approx 6$ (implying a short-wavelength cutoff) while, for $\Gamma=200$, the spectrum becomes very
noisy beyond $ka\approx 6$, so that it is difficult to distinguish peaks due to collective motion. One can also notice that, for
$\Gamma=100$, the peak contour of the wave spectra does not seem to reach $ka=0$ at $\omega=0$, indicating a long-wavelength
cutoff \cite{Murillo2000}. (This feature will be further discussed below.) Agreement between the simulation and the QLCA does
not appear to be very satisfactory, neither at short nor at long wavelengths. The agreement at long wavelengths is somewhat
improved for $\Gamma=200$, in which case the long wavelength cutoff, implied by the simulation, vanishes. It should be noted
that the melting point for this system occurs at $\Gamma^{*}\approx 180$ for $\kappa=1$ \cite{Kalman2004}. So, at $\Gamma=200$,
particles are almost frozen and wave propagation becomes anisotropic at short wave lengths. The task of resolving angular
dependence of the wave dispersion lies beyond the capability of QLCA, but this issue can be tackled by the HA, as shown later
on.

\subsection{QLCA with DTW in longitudinal mode}
\begin{figure}[htp]
\centering
\includegraphics[trim=5mm 4mm 5mm 0mm,clip, width=0.75\textwidth]{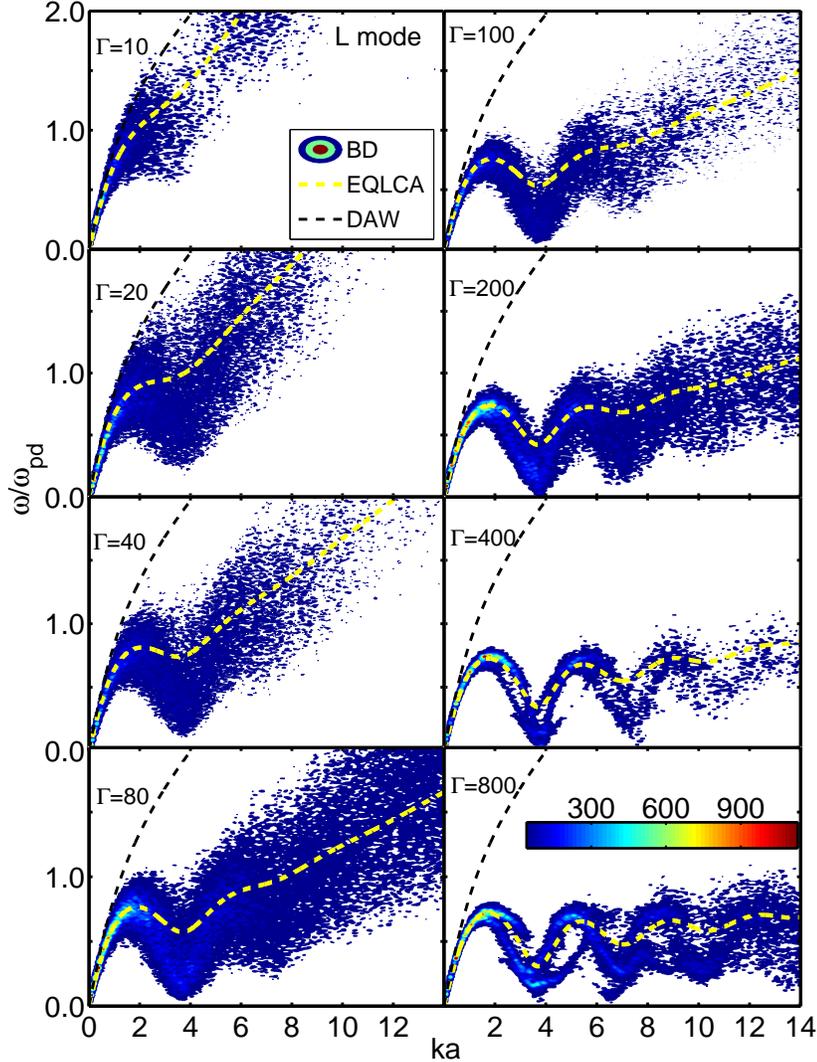}
\caption{(Color online) (Color bar in arbitrary unit) Wave spectra of longitudinal mode for $\kappa=1$ and $\nu=0.01\omega_{pd}$ but different $\Gamma$ covering both liquid
and solid states. Thin dashed lines are results of DAW, while the heavy dashed lines are extensions of QLCA with DTW modes, i. e.,
EQLCA} \label{fig_L_diff_G}
\end{figure}

It is known \cite{Golden2000,Kalman2004,Donko2008} that the QLCA neglects the direct thermal effect, which is responsible for the actual
motions and migration of dust particles, and which gives rise to the so-called Bohm-Gross term ($\propto k^2v^{2}_{th}$) in the
third-frequency-momentum sum rule, and consequently in the longitudinal dispersion relation \cite{Golden2000}. Since this term
is of the order of $O(\Gamma^{-1})$ in comparison with terms due to the correlation effect, it is a good approximation to
neglect it for $\Gamma>>1$ and, especially, at long wavelengths, $ka<<1$. At short wavelengths, $ka>>1$, the $k^{2}$ factor
could elevate the direct thermal effect up to a significant level. Nevertheless, this effect was not examined in previous
simulations, and is therefore of interest for the present study.

A simple extension of the QLCA, which would include the effect of direct thermal motion, can be made at the phenomenological level by
taking into account a contribution from the kinetic part of pressure. Mathematically, this can be done by inserting the QLCA-LFC
into Eq. (\ref{Eq_dispersion_DTW}), in which case it gives a correction to the longitudinal mode from the QLCA dielectric
function, as follows
\begin{equation}
\varepsilon\left( k,\omega \right) =1-\frac{1}{\omega(\omega+i\nu)-D_{L}(k)}\left [\omega_{0}^2(k)
+\frac{\gamma_{d}v^{2}_{th}}{2}k^{2} \right ], \label{Eq_dielectric_EQLCA}
\end{equation}
and consequently the longitudinal dispersion relation becomes
\begin{equation}
\omega(\omega+i\nu) =\omega^{2}_{0}(k)+D_L(k)+\frac{\gamma_{d}v^{2}_{th}}{2}k^{2} .  \label{Eq_dispersion_EQLCA}
\end{equation}

Figure \ref{fig_L_diff_G} shows wave spectra of the longitudinal mode for different values of $\Gamma$ covering a wide range in
both liquid and solid states. Dispersion relations of both the DAW and the QLCA with the above DTW extension (denoted as EQLCA), i. e., Eq.\ref{Eq_dispersion_EQLCA},
are also shown for comparison. It is seen that the raising DTW tails in the spectra become less and less significant with
increasing $\Gamma$. The DTW mode is still noticeable even at $\Gamma=800$, although it is very weak there, as shown. On the
other hand, one also sees that the EQLCA captures the DTW tails in the spectra, in addition to exhibiting a consistently good
agreement with simulation at small wave-numbers. Nevertheless, discrepancies still exist in some fine structures at intermediate
wavelengths. Some of those, occurring for $\Gamma>\Gamma^{*}$, are due to the angular-dependence of wave propagation in the
solid state.

It should be noted here that our extension is essentially based on mean-field theory of Rao\cite{Rao2000} and that QLCA is used only as an LFC. Therefore from this point of view, our extension is not strictly self-consistent and might not satisfy the third frequency sum rule. A more self-consistent theory, which can satisfy the third frequency sum rule, is now under developing and will be presented elsewhere \cite{Murillo2009}.

\subsection{QLCA with critical wave cutoff in transverse mode}
\begin{figure}[htp]
\centering
\includegraphics[trim=15mm 14mm 15mm 0mm,clip, width=0.7\textwidth]{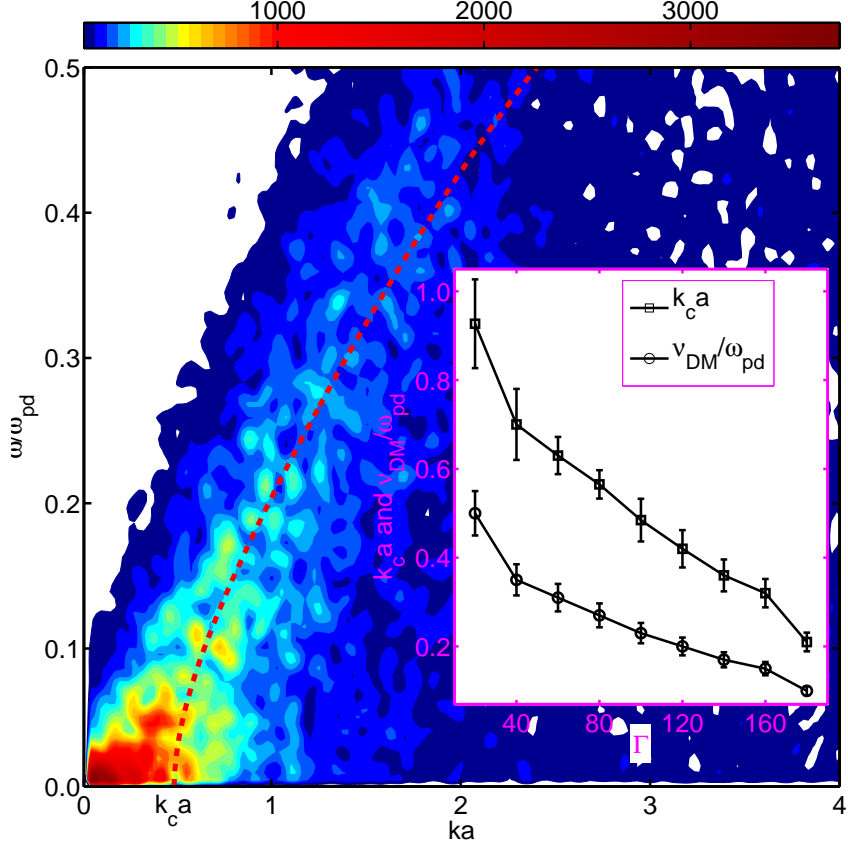}
\caption{(Color online) (Color bar in arbitrary unit) Transverse mode for $\Gamma=100$, $\kappa=1$ and $\nu=0.01\omega_{pd}$. Dashed line is the fit according to Eq.
(\ref{Eq_EQLCA_T1}), i. e., the extension of QLCA with cutoff wave-number. This gives a $\nu_{DM}\approx 0.23\omega_{pd}$ and a
cutoff wave-number of $k_{c}a=0.48$, as is labeled in the figure. The inserted figure shows $k_{c}a$ for $\kappa=1$,
$\nu=0.01\omega_{pd}$ but different $\Gamma$.} \label{fig_T_K1G200}
\end{figure}

An another weakness of the QLCA lies in its inability to account for the diffusional and other damping effects that preclude the
existence of long wavelength transverse waves in the liquid state \cite{Kalman2004}. Kalman \emph{et al.} \cite{Kalman2004}
provided a work-around by introducing a phenomenological damping $\nu_{DM}$. This gives rise to a small change in the transverse
dielectric function of the QLCA,
\begin{equation}
\varepsilon\left( k,\omega \right) =1-\frac{\omega_0^2(k)}{\omega(\omega+i\nu+i\nu_{DM})-D_{T}(k)}, \label{Eq_EQLCA_T}
\end{equation}
and consequently in the transverse dispersion relation,
\begin{equation}
\omega(\omega+i\nu+i\nu_{DM})=D_T(k). \label{Eq_disp_EQLCA_T}
\end{equation}
By assuming $\nu=0^{+}$, the real part of the dispersion relation can be written as
\begin{equation}
\omega_{r} = \sqrt{D_T(k)-\frac{\nu^{2}_{DM}}{4}}, \label{Eq_EQLCA_T1}
\end{equation}
so that the condition $D_T(k)-\nu^{2}_{DM}/(4)\ge 0$ gives rise to a critical cutoff wave-number $k_{c}$.

Note that $\nu_{DM}$, and consequently $k_{c}$, can be determined by directly fitting the simulation results to the above form
\cite{Kalman2004}. Results of such fitting are shown in Fig.\ \ref{fig_T_K1G200}, along with a magnification of Fig.\
\ref{fig_T_K1G100} at small $k$s, where one can clearly notice the existence of a long wavelength cutoff $k_{c}$. As a
consequence, one can notice an improved agreement between the simulation and the QLCA with the extension introducing the cutoff.
This extension of the QLCA does not affect much the dispersion at large $k$, where it behaves in much the same way as in Fig.
\ref{fig_T_K1G100} because $\nu_{DM}\approx 0.23\omega_{pd}$ is relatively small. From the figure's inset, one sees that the
cutoff wave-number $k_{c}$ increases with decreasing $\Gamma$. This tendency agrees with previous experimental observations
\cite{Nosenko2006}.

\subsection{HA for DLW}
\begin{figure}[htp]
\centering
\includegraphics[trim=15mm 2mm 12mm 0mm,clip, width=0.8\textwidth]{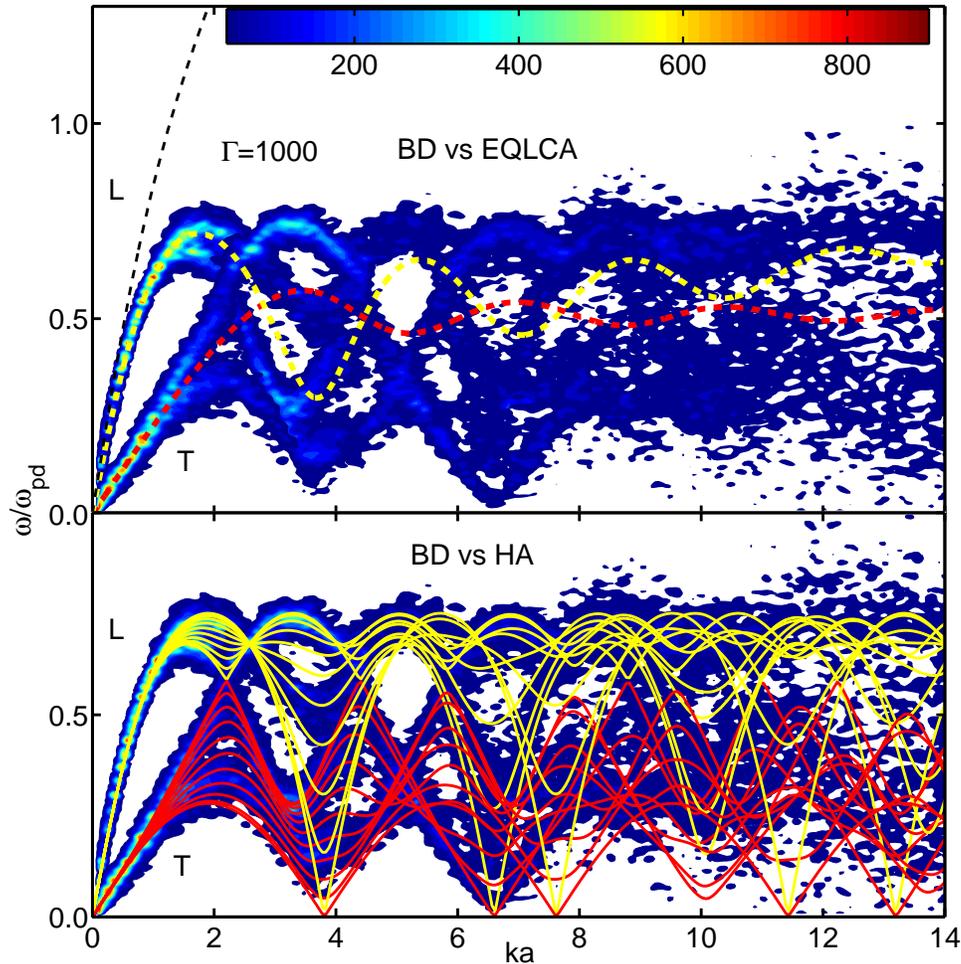}
\caption{(Color online) (Color bar in arbitrary unit) Wave spectra of both longitudinal and transverse modes for $\kappa=1$, $\Gamma=1000$ and $\nu=0.01\omega_{pd}$. The
upper panel shows the result of BD versus EQLCA and also DAW, while the lower panel for BD versus HA. The results of HA cover a
whole period of $\pi/6$ with an interval of $3$ degree.} \label{fig_LT_HA}
\end{figure}

In a perfect crystalline state, all particles are located at the points of a triangular lattice and they perform thermal
oscillations around their equilibrium positions. In this case, one needs to adopt an another strategy to obtain dispersion
relation for the DLW. This dispersion relation can be determined from the following eigenvalue problem \cite{Dubin2000}:
\begin{equation}
\lVert \omega^{2}(\mathbf{k})\mathbf{I}-\mathbf{M}(\mathbf{k}) \rVert=0,\label{Eq_M}
\end{equation}
where $\mathbf{I}$ is the 2D unit matrix. (Note that, although we have neglected the phenomenological damping in Eq.
(\ref{Eq_M}), it can be easily retained by replacing $\omega^{2}$ with $\omega(\omega+i\gamma)$.) The matrix
$\mathbf{M}(\mathbf{k})$ is the interaction matrix, given by \cite{Dubin2000}
\begin{equation}
\mathbf{M}(\mathbf{k})=\frac{1}{m}\sum_{i}\frac{\partial^{2}\phi}{\partial\mathbf{r}_{i}\partial\mathbf{r}_{i}}[1-e^{i\mathbf{k\cdot}\mathbf{r}_{i}}]\equiv
[M_{\alpha\beta}(\mathbf{k})], \label{Eq_M_HA}
\end{equation}
where $\alpha$ and $\beta=x,y$, and the summation over $i$ includes all points on the triangular lattice.

For the HA \cite{Dubin2000}, results are quite similar to those of QLCA. There are two branches given by,
\begin{eqnarray}
\omega^2_{L/T}(k,\theta)&=&\frac{1}{2}\left [ M_{xx}+M_{yy} \right ] \nonumber \\
&\pm & \frac{1}{2}\sqrt{[M_{xx}+M_{yy}]^2+4(M^{2}_{xy}-M_{xx}M_{yy})}, \label{Eq_wk_HA}
\end{eqnarray}
where $M_{\alpha\beta}$ is given by Eq. (\ref{Eq_M_HA}), and $\theta$ is the polarization angle \cite{Dubin2000}. The subscripts
${L}$ and ${T}$ denote the longitudinal and transverse modes, and they correspond to $+$ and $-$ signs in Eq. (\ref{Eq_wk_HA}),
respectively. Therefore, wave propagation depends on the angle $\theta$, and its dispersion relations are functions with the
period of $\pi/6$ due to the hexagonal symmetry \cite{Dubin2000}. The angular dependence arises because the system is now
anisotropic at short wavelengths.

Figure \ref{fig_LT_HA} shows wave spectra of both longitudinal and transverse modes in a crystalline state with $\Gamma=1000$.
Also shown are the dispersion relations of the HA, EQLCA and DAW. The HA curves cover the whole period of $\pi/6$ with an
increment of $3$ degrees. It is seen that the simulation agrees very well with the HA, and that the polarization effects in the
simulation spectra are fully captured by the HA. Good agreement between the EQLCA and the simulation is retained in the first
Brillouin zone. For large wave-numbers, the EQLCA cannot resolve the polarization, i.e., the angular-dependence of the
dispersion. Nevertheless, it captures the correct tendency for oscillations. Moreover, at this range of $\Gamma$ values, the
QLCA regains its credibility because both the thermal effect and the long wavelength cutoff corrections become unimportant, and
the EQLCA essentially reduces to the standard QLCA. At the same time, it is remarkable to see that the mean field DAW is so
robust, even in a crystalline state, where it still offers a fairly good agreement with the simulation at long wavelengths.
\begin{figure}[htp]
\centering
\includegraphics[trim=5mm 2mm 12mm 0mm,clip, width=0.7\textwidth]{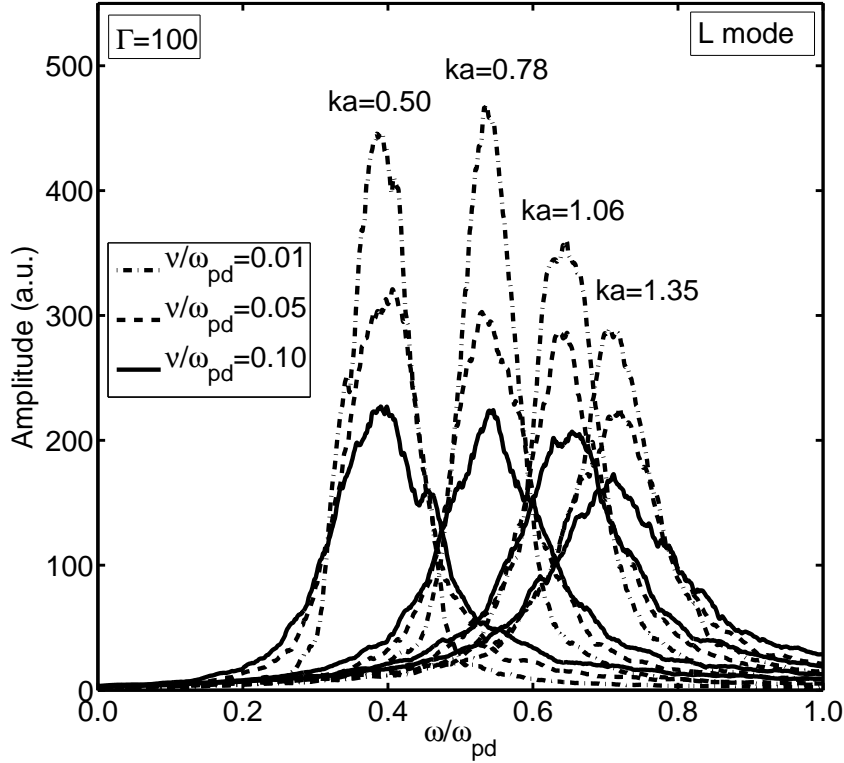}
\caption{Profiles of longitudinal wave spectra for certain wavenumbers with $\kappa=1$, $\Gamma=100$ but different damping rate
$\nu$.} \label{fig_L_damping}
\end{figure}

\subsection{Damping effect}

\begin{figure}[htp]
\centering
\includegraphics[trim=15mm 5mm 12mm 5mm,clip, width=0.8\textwidth]{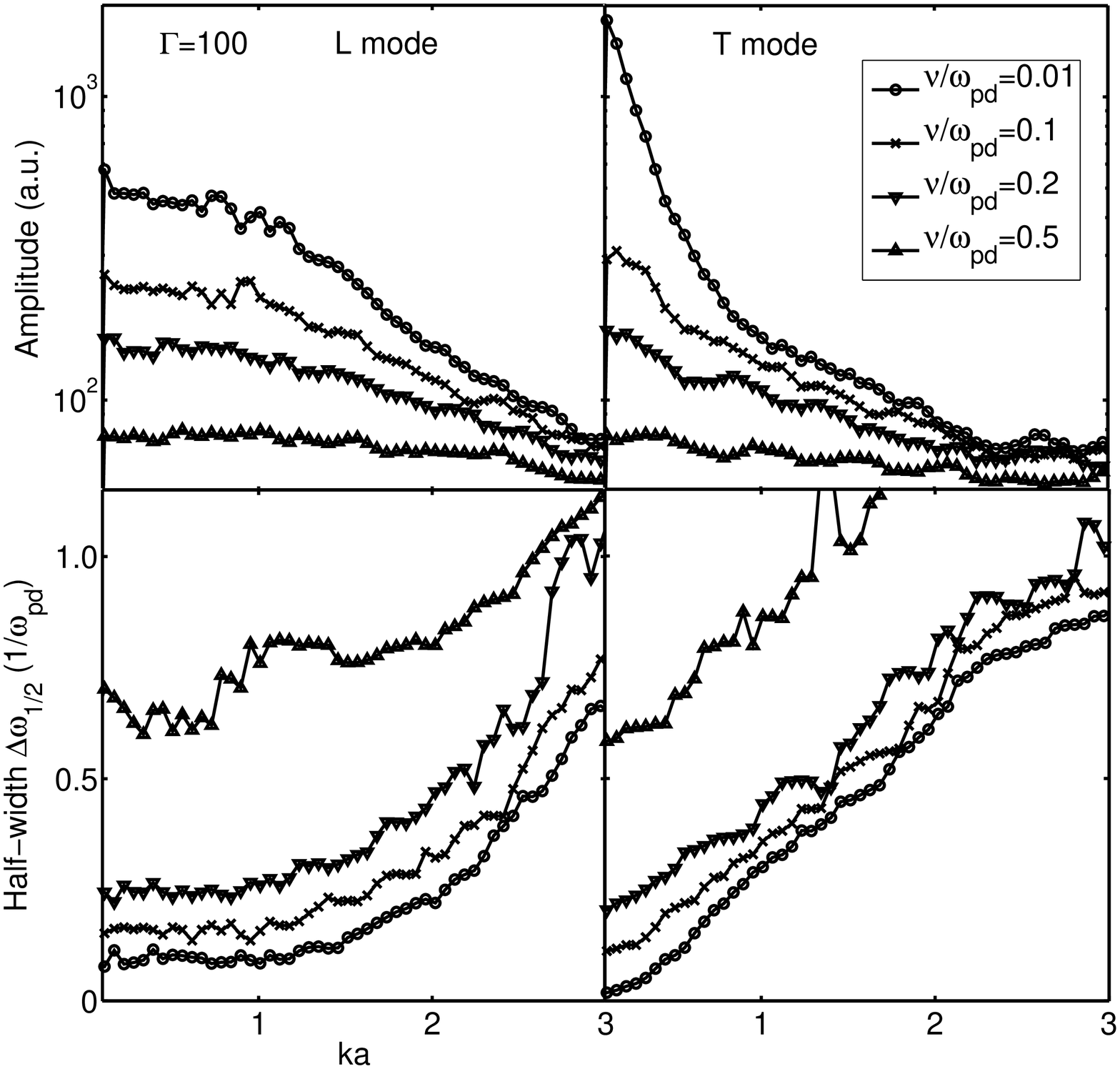}
\caption{Peak-amplitudes and half-widths $\Delta\omega_{1/2}$ of both longitudinal and transverse wave spectra for $\kappa=1$,
$\Gamma=100$ but different neutral gas damping rate $\nu$.} \label{fig_LT_damping}
\end{figure}

We finally investigate the effects of damping on wave spectra. Generally speaking, there are two types of damping mechanisms for
collective modes, i.e., the intrinsic and external ones. The former type includes damping due to diffusion, viscosity, thermal
conduction and, possibly, also Landau-like damping \cite{Boon1980,Ichimaru,Golden2000}, while the latter type mainly comes from
the neutral gas. The external damping is simply given by the imaginary part of the dispersion relation by $\omega_{i}=-\nu/2$, , whereas the intrinsic damping of collective modes involves complicated physical processes, and so far there
has been no well founded theory of such processes for SCDPs. Development of such a theory may be an interesting topic for future
research, while, at this stage, we present several results from our simulation.

Figure \ref{fig_L_damping} shows the profiles, or, more precisely, the cross sections at several fixed wave numbers, of the
longitudinal wave spectra, $\mathcal{L}(\mathbf{k},\omega)$, for $\Gamma=100$ and with different neutral damping rates $\nu$.
These profiles exhibit typical resonance shapes which describe the spectral distributions shown in Figs.\ 1-7, while their peak
positions in the $\omega$-$k$ plane correspond to the dispersion relations. The above mentioned analytical theories give only
the peak positions in the $\omega$-$k$ plane. In reality, these peaks are never delta-like functions, but rather have finite
amplitudes and consequently finite widths due to various damping effects, as is shown in Fig.\ \ref{fig_L_damping}. Physically,
the peak amplitudes indicate the amount of energy concentrated in various modes, while the widths of the peaks at half heights
(denoted as the half-widths, $\Delta\omega_{1/2}$) are related to damping \cite{Nunomura2002}. One notices a general tendency
that the peak amplitudes decrease with the increase of wave-number. Similar tendency was also observed in classical MD simulation \cite{Donko2008}. In addition, with the increase of the neutral gas damping
rate $\nu$, all the heights decrease, while the widths increase. These tendencies are depicted more clearly in Fig.
\ref{fig_LT_damping}, which shows the amplitudes and the half-widths of $\mathcal{L}(\mathbf{k},\omega)$ and
$\mathcal{T}(\mathbf{k},\omega)$ versus wave-number for different $\nu$.

\section{Conclusion}
By using the Brownian dynamics simulation, we have investigated the wave spectra of 2D dusty plasmas in states that cover a full
range from a non-ideal gas to crystalline. The results are critically compared with dispersion relations for the dust-acoustic
wave, dust-thermal wave, and those from the quasi-localized charge approximation and the harmonic approximation. In particular,
simply extensions are considered to include the direct thermal effect and the critical cutoff wavenumber into the QLCA. It is found
that, for a non-ideal gaseous state (e.g., with $\Gamma=1$), the longitudinal DTW is in a remarkably good agreement with the
simulation, while for a perfect crystalline state (e.g., $\Gamma=1000$), the HA agrees very well with the simulation. In the
states between those two opposing ends, an overall good agreement between the extended versions of the QLCA and the simulation
was found for a wide range of wavelengths. The damping effect is also briefly discussed in terms of the peak amplitudes and the
half-widths of the current-current correlation function for different neutral gas damping rate $\nu$. A general tendency is
that the peak amplitudes decrease with the increase of wave-number and the increase of $\nu$, while the widths increase during these courses.

\begin{acknowledgments}
L.J.H. acknowledges support from Alexander von Humboldt Foundation. Work at CAU is supported by DFG within
SFB-TR24/A2. Z.L.M. acknowledges support from NSERC. L.J.H thanks Prof. Z. Donk\'{o} and Prof. P. K. Shukla for valuable discussions.
\end{acknowledgments}

\end{document}